\documentclass[11pt]{article}
\usepackage{graphicx}
\usepackage{times}
\usepackage[hmargin=3cm,bottom=3.5cm,top=2.5cm, bindingoffset=0.5cm]{geometry}
\usepackage{epstopdf}
\usepackage{amsmath}
\usepackage{hyperref}
\usepackage{xcolor}
\hypersetup{colorlinks=true, linkcolor=black, citecolor=black}
\usepackage[english]{babel}
\usepackage[numbers,sort&compress]{natbib}
\usepackage{authblk}
\usepackage{caption}
\usepackage[font=scriptsize]{caption}
\title{Robust stabilization loop design for gimbaled electro-optical imaging system}

\date{}
\author[1]{Mehmet Bask{\i}n}
\affil [1]{Aselsan Inc., Ankara, Turkey}
\author[2]{Kemal Leblebicio{\u{g}}lu}
\affil [2]{Dept. of EEE, Middle East Technical University, Ankara, Turkey}
\begin{document}
\maketitle
\begin{abstract}
For electro-optical imaging systems, line-of-sight stabilization against different disturbances created by mobile platforms is crucial property. The development of high resolution sensors and the demand in increased operating distances have recently increased the expectations from stabilization loops. For that reason, higher gains and larger bandwidths become necessary. As the stabilization loop satisfies these requirements for good disturbance attenuation, it must also satisfy sufficient loop stability. In gimbaled imaging systems, the main difficulties in satisfying sufficient loop stability are structural resonances and model uncertainties. Therefore, satisfying high stabilization performance in the presence of model uncertainties or modeling errors requires utilization of robust control methods. In this paper, robust LQG/LTR controller design is described for a two-axis gimbal. First, the classical LQG/LTR method is modified such that it becomes very powerful loop shaping method. Next, using this method, controller is synthesized. Robust stability and robust performance of stabilization loop is investigated by using singular value tests. The report is concluded with the experimental validation of the designed robust controller.\\

\noindent {\bf keywords:} LQG/LTR, robust multivariable control, line-of-sight stabilization, multi-axis gimbal, loop shaping
\end{abstract}
\section{Introduction} 
For precise pointing and tracking performance, line-of-sight (LOS) stabilization against various disturbances is essential for imaging systems. This problem is usually solved by using a high performance stabilization loop. High stabilization performance requires high gains and large bandwidths. As the stabilization loop posses these properties for good disturbance attenuation, it must also satisfy sufficient loop stability. In gimbaled imaging systems, the main difficulties in satisfying sufficient loop stability are structural resonances and model uncertainties. Therefore, satisfying high stabilization performance in the presence of model uncertainties or modeling errors requires utilization of robust control methods. In this aspect, this paper is devoted to the design of a stabilization loop for a two-axis gimbal.\\
\indent Classical controller with PI and lead lag compensators was preferred in stabilization loops in the past \cite{kennedy2013line,masten2008inertially}. However, finding the classical controller that satisfies both stability and performance criteria is time consuming iterative procedure. Moreover, this method is insufficient in optimality aspect. Recently, different techniques are used for better stabilization performance. Linear quadratic methods are explained in \cite{bigley1987wideband,marathe1997lqg,lee1998scan}. Moreover, the H$_\infty$ control methods are discussed in \cite{lee1998scan,moorty2002h,liu2013h}. However, in most of the designs, analysis of performance change under model perturbations is missing. In gimbaled systems, structural resonances, sensor delays and nonlinear friction are typical source of the model
perturbations. For that reason, the stabilization loop must be robust to satisfy good performance in the presence of model uncertainty. On the other hand, the robust control methods in \cite{lee1998scan,wenliang2010control,kim2010robust} are not supported with theoretical and experimental data to validate the robustness of closed loops.\\
\indent In previous LQG/LTR designs, desired loop shape is obtained by adjusting the
weighting matrices or intensities of process and measurement noises. However in this
paper, the design is modified such that it becomes a powerful loop shaping method \cite{maciejowski1989multivariable}. In other words, the sensitivity at the plant output is successfully shaped for good disturbance rejection. After designing the controller, the robustness of the design is investigated by
using theoretical results. Finally, the theoretical results are supported with experimental data
to validate the robustness of the stabilization loop.
\section{LQG/LTR control} \label{s2}
Traditional LQG control method assumes that the plant is linear time invariant, measurement and process noises are stochastic with known statistical properties \cite{maciejowski1989multivariable}. The
plant is represented with a state space representation in \eqref{a1},
\begin{align} \label{a1}
\begin{split}
\dot{x}&=A x+B u+\Gamma w_{d}\\
y&=C x + w_{n}
\end{split}
\end{align}
where $ w_{d} $ and $ w_{n} $ are uncorrelated zero mean white noise processes having constant power spectral densities $ W $ and $ V $ as illustrated in \eqref{a2}.
\begin{align} \label{a2}
\begin{split}
E \left\{ w_{d} (t) w_d^T (\tau) \right\} = W\delta (t-\tau),~
E \left\{ w_{n} (t) w_n^T (\tau) \right\} = V\delta (t-\tau),~
E  \left\{ w_{d} (t) w_n^T (\tau) \right\} = 0
\end{split}
\end{align}\\
The objective of the LQG theory is to minimize the cost function given in \eqref{a3} where
$ Q = Q^T \geq 0 $ and $ R = R^T > 0 $ are weighting matrices. The solution of this problem can be obtained in two steps:
\begin{equation} \label{a3}
\ J= \lim_{T \to \infty } E \left\{   \int_{0}^{T} (x^TQx+u^TRu)dt \right\}
 \end{equation}  
\indent Step 1: Obtain an optimal estimate $\hat{x}$ of states $x$ such that $ E \left\{(x-\hat{x})^T (x- \hat{x}) \right\} $ is minimized.\\
\indent Step 2: Use estimate as if it were true state measurement and solve LQ regulator problem.\\
\indent The solutions of these two problems, Kalman filter and LQ regulator, both have very
good stability properties individually. It is reported in \cite{maciejowski1989multivariable,anderson2007optimal} that Kalman filter and LQ regulator can tolerate gain variation between $(1/2, \infty)$ and phase variation less than $60^{\circ}$ in each channel. However when they are combined, there is no guaranteed stability conditions for LQG regulators. Moreover, LQG regulators may suffer from poor stability margins if the designers do not pay enough attention \cite{doyle1978guaranteed}. \\
\indent Loop transfer recovery method introduced in \cite{doyle1981multivariable} overcomes this drawback of LQG regulators. In this method, optimal state feedback is designed such that the Kalman filter properties are recovered at the plant output. The procedure can be summarized as below where the notation  $\Phi= (sI-A)^{-1} $ is used:\\
\indent Step 1: By adjusting the covariance matrices $W$ and $V$, design a Kalman filter such that
the desired open loop transfer matrix $C\Phi K_f$ is obtained.\\
\indent Step 2: Design a LQ regulator by choosing $Q=I$ and $R= \rho I$, and reduce $\rho$ until the open
loop transfer matrix at the plant output approaches enough to $C\Phi K_f$ over necessary
frequency interval \cite{maciejowski1989multivariable}.\\
\indent Obtaining a good Kalman filter open loop $C\Phi K_f$ is not an easy task. Now, very effective and simple procedure that gives a good Kalman filter shape will be discussed.
\subsection{Shaping singular values}
\begin{figure}[!hbt] 
	\centering
	\captionbox	
	{Plant augmentation \label{fi1}}
	{\includegraphics[scale=0.75]{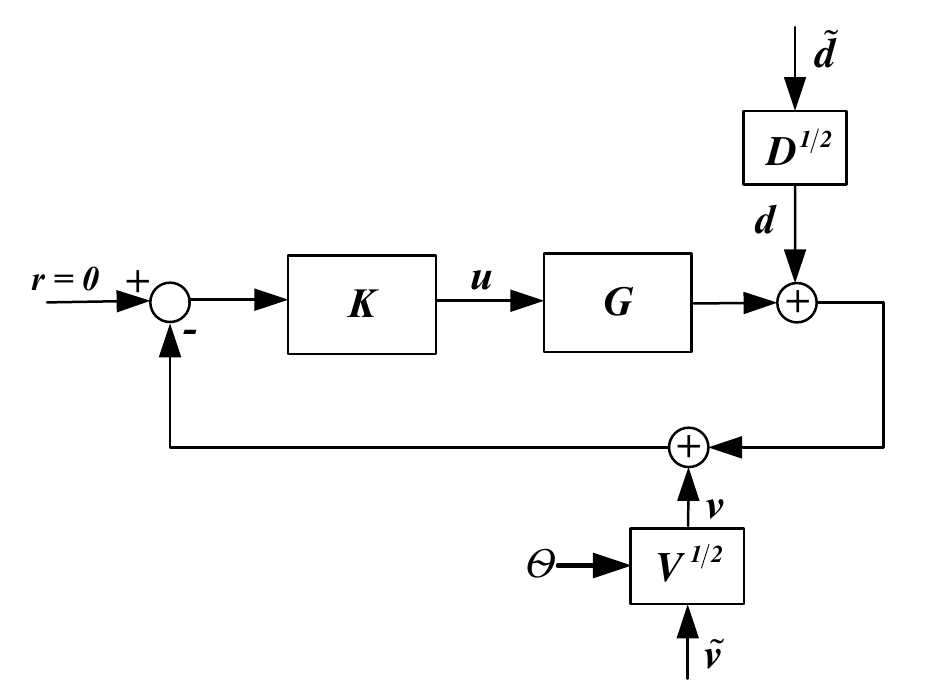} }
\end{figure}

To design a satisfactory Kalman filter open loop, designer should modify $W$ and $V$.
However, if frequency dependent weighting matrices $W(s)$ and $V(s)$ are used, to obtain a good $C\Phi K_f$ is simpler. In this work, this powerful and simple loop shaping technique
reported in \cite{stein1987lqg} is used.\\
\indent Assume that as in Figure \ref{fi1}, instead of state disturbances the plant has a disturbance $d$ having power spectral density $D(s)$ and measurement noise $v$ having power spectral density $V (s)$. Moreover let the disturbance $d$ and measurement noise $v$ are created from the processes \eqref{a4} and \eqref{a5} respectively,
\begin{align} \label{a4}
\begin{split}
\dot{\xi}&=A_d \xi +B_d \tilde{d}\\
d&=C_d\xi
\end{split}
\end{align}
\begin{align} \label{a5}
\begin{split}
\dot{\eta}&=A_v\eta +B_v \tilde{v}\\
v&=C_v\eta+\Theta
\end{split}
\end{align}
where $\tilde{d}$, $\tilde{v}$ and $\Theta$ are white noise processes. If one combines the states of the original plant and these two processes, augmented system shown in \eqref{a6} is obtained.
\begin{align} \label {a6}
\begin{split}
\begin{bmatrix}
\dot{x}  \\ \dot{\xi}  \\ \dot{\eta} \end{bmatrix}=&
\begin{bmatrix}
A & 0 & 0 \\
0 & A_d & 0 \\
0 & 0 & A_v \end{bmatrix}
\begin{bmatrix} x \\ \xi  \\ \eta \end{bmatrix}+
\begin{bmatrix} B \\ 0  \\ 0 \end{bmatrix}u+
\begin{bmatrix} 0 & 0  \\ B_d & 0 \\ 0 & B_v \end{bmatrix} 
\begin{bmatrix} \tilde{d} \\ \tilde{v} \end{bmatrix} \\
y=&\begin{bmatrix} C & C_d & C_v \end{bmatrix}
\begin{bmatrix} x \\ \xi  \\ \eta \end{bmatrix}+\Theta \\
z=&\begin{bmatrix} C & C_d & C_v \end{bmatrix}
\begin{bmatrix} x \\ \xi  \\ \eta \end{bmatrix} \\
\end{split}
\end{align}
The modified plant can still be used in LQG framework by assuming \eqref{a7} is satisfied. In other words, LQG compensator can be designed for this augmented bigger plant in \eqref{a6}.
\begin{align} \label{a7}
\begin{split}
E\left\{ \Theta(t) \Theta^T (\tau) \right\} > 0,~
E \left\{ \Theta(t)  \tilde{v}^{T} (\tau) \right\} =0,~
E  \left\{ \Theta(t)   \tilde{d}^{T} (\tau) \right\} =0
\end{split}
\end{align}
For the structure in Figure \ref{fi1}, the closed loop equations \eqref{a8} and \eqref{a9} are used.
\begin{equation}\label{a8} z=S_{o}d-T_{o}v \end{equation}
\begin{equation}\label{a9} u=-KS_{o}d-KS_{o}v \end{equation}
If the designer apply the LTR procedure for augmented plant, the cost of the LTR procedure approaches \eqref{a10} by taking $Q=I$, $R=\rho I$ and by reducing $\rho$ \cite{stein1987lqg}.
\begin{equation} \label{a10}
\lim_{\rho \to 0 } J_{LTR}=\frac{1}{2\pi} \left\{   \int_{-\infty}^{\infty} \sum_{i=1}\sigma^{2} [S_{o} D^{1/2}(jw)]+
\sum_{i=1}\sigma^{2} [T_{o} V^{1/2}(jw)]dw \right\}
\end{equation}  
It can be seen that LTR procedure applied at plant output trades off the output sensitivity
$S_{o}(jw)$ against the output complementary sensitivity $T_{o} (jw)$ with a factor
$W_{e} (jw)=D^{1/2} (jw) V^{-1/2} (jw)$. 
After assuming $V=I$ and choosing $D^{1/2} (jw)$  appropriately, it is
possible to shape the sensitivity function over required frequency ranges for good disturbance rejection \cite{maciejowski1989multivariable}.
\begin{figure}[!hbt]
	\centering	
	\captionbox	
	{Gimbal model for one axis \label{fi2}}
	{\includegraphics[scale=0.75]{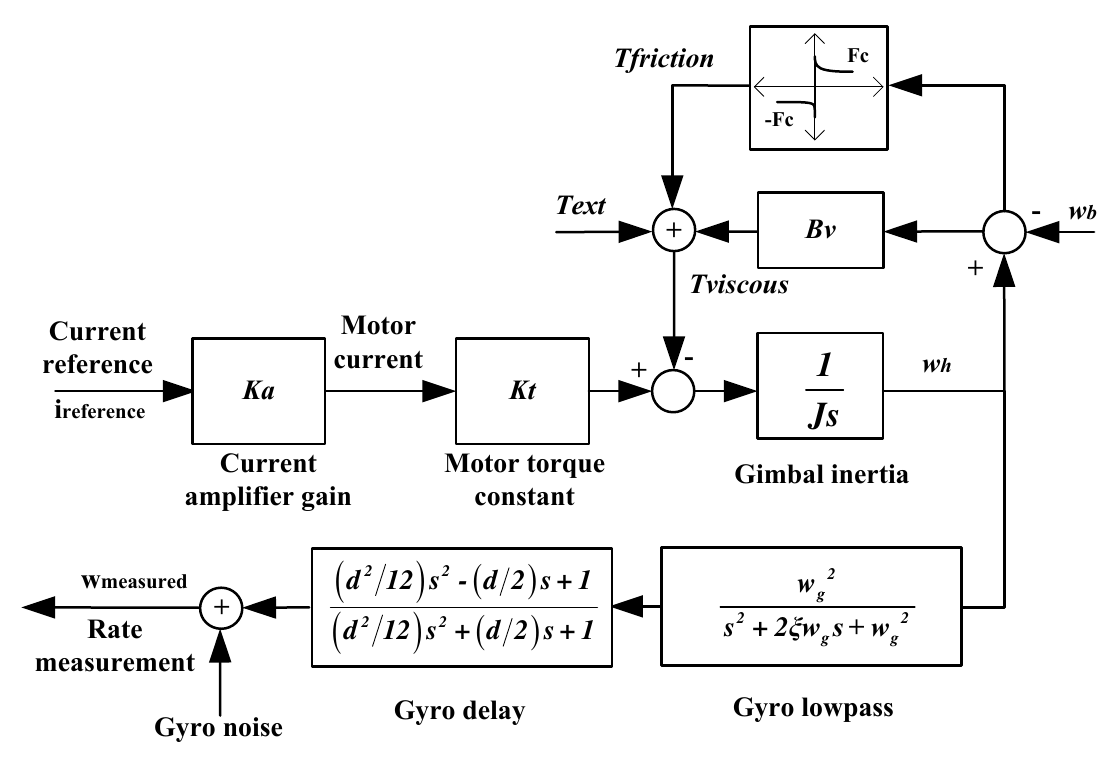}}
\end{figure}
\section{Two-axis gimbal model}
The dynamic equations of the azimuth-elevation gimbal show that when the gimbal is
designed to be mass balanced, the azimuth and elevation equations decouple \cite{baskin2015lqg,ekstrand2001equations}. In
other words, the angular rate of any axis depends only on the net torque applied to that axis,
and for one axis model given in Figure \ref{fi2} can be used. By assuming that the dynamic
friction only depends on inertial rate and neglecting the static friction, the model used for
stabilization is approximated as in \eqref{a11} \cite{baskin2015lqg}.
\begin{equation} \label{a11}
	G(s)=\frac{w_{measured}}{i_{reference}}=\frac{K_a K_t}
	{Js+B_v} \times \frac{w_g^2}{s^2+2\xi w_g s+w_g^2} \times \frac{(d^2/12)s^2-(d/2)s +1}{(d^2/12)s^2+(d/2)s +1}	
\end{equation}
From the datasheets of the motor, driver and gyro, the parameters listed in Table \ref{t1} are
obtained. According to model \eqref{a11}, inertia $J$ and viscous constant $B_v$ need to be found.
Determination of inertia $J$ and viscous constant $B_v$ is not easy task and it requires more complicated analysis.
\subsection{Extended Kalman filter for parameter estimation}
The parameter identification method through state augmentation is nonlinear, and
nonlinear filtering technique needs to be utilized. In this paper, extended Kalman filter
(EKF) which is the most common nonlinear filtering technique is used for unknown parameters estimation.
\begin{table}
	\caption{Parameters of the system \label{t1}} 
	\begin {center}
	\begin{tabular}{l l}
		\hline
		Parameters & Values \\
		\hline
		current aplifier gain, $K_a$ & $2~A/A$ \\
		motor torque constant, $K_t$ & $2.18~Nm/A$ \\
		natural frequency of rate gyro, $w_g$ & $1646~rad/s$ \\
		damping of gyro, $\xi$ & $0.8$ \\
		gyro delay, $d$ & $4.5~ms$ \\
		\hline
	\end{tabular}
\end{center}
\end{table}
\subsubsection{Problem simplification}
\begin{figure}[!hbt]
 	\centering	
 	\captionbox	
 	{CD-EKF results \\ a Inertia estimation \\b Viscous friction constant estimation \label{fi4}}
 	{\includegraphics[scale=0.7]{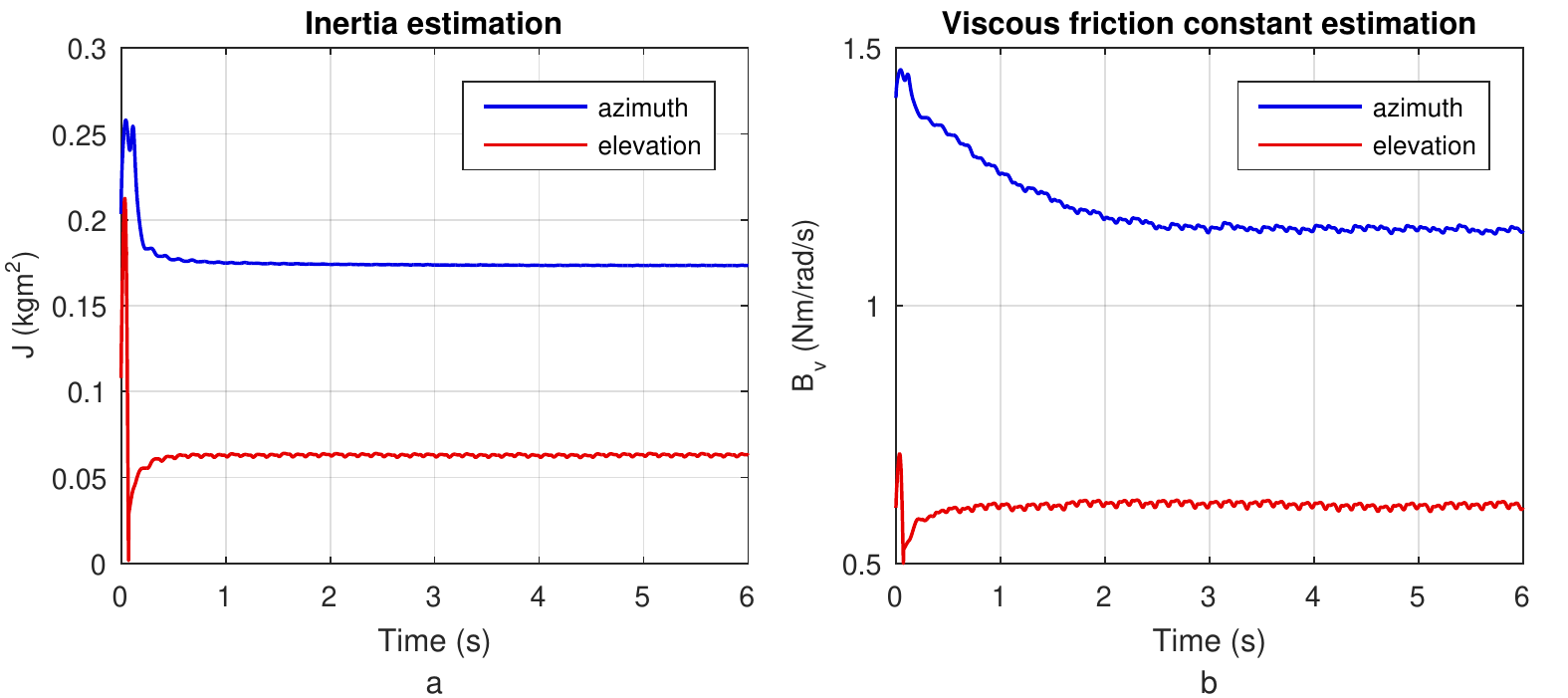}}
 \end{figure}

While using continuous-discrete extended Kalman filter (CD-EKF), at the time update stage, the states and entries of covariance matrices are found solving differential equations. In this aspect, to solve the parameter identification problem, it is necessary to keep the model as simple as possible. To get rid of the singularity problems in numerical solution of differential equations, the delay is approximated with a first order low pass filter as in \eqref{a12}. This assumption is only valid when the system is excited with a low frequency signal where
the magnitude and phase responses of these two transfer functions are very close.
\begin{equation} \label{a12}
	\frac{(d^2/12)s^2-(d/2)s +1}{(d^2/12)s^2+(d/2)s +1}\approx\frac{1}{ds+1}
\end{equation}
\begin{figure}[!hbt] 
	\centering	
	\captionbox	
	{Pade approximation and first order low pass \\
		a Magnitude plot \\ b Phase plot \label{fi3}}
	{\includegraphics[scale=0.7]{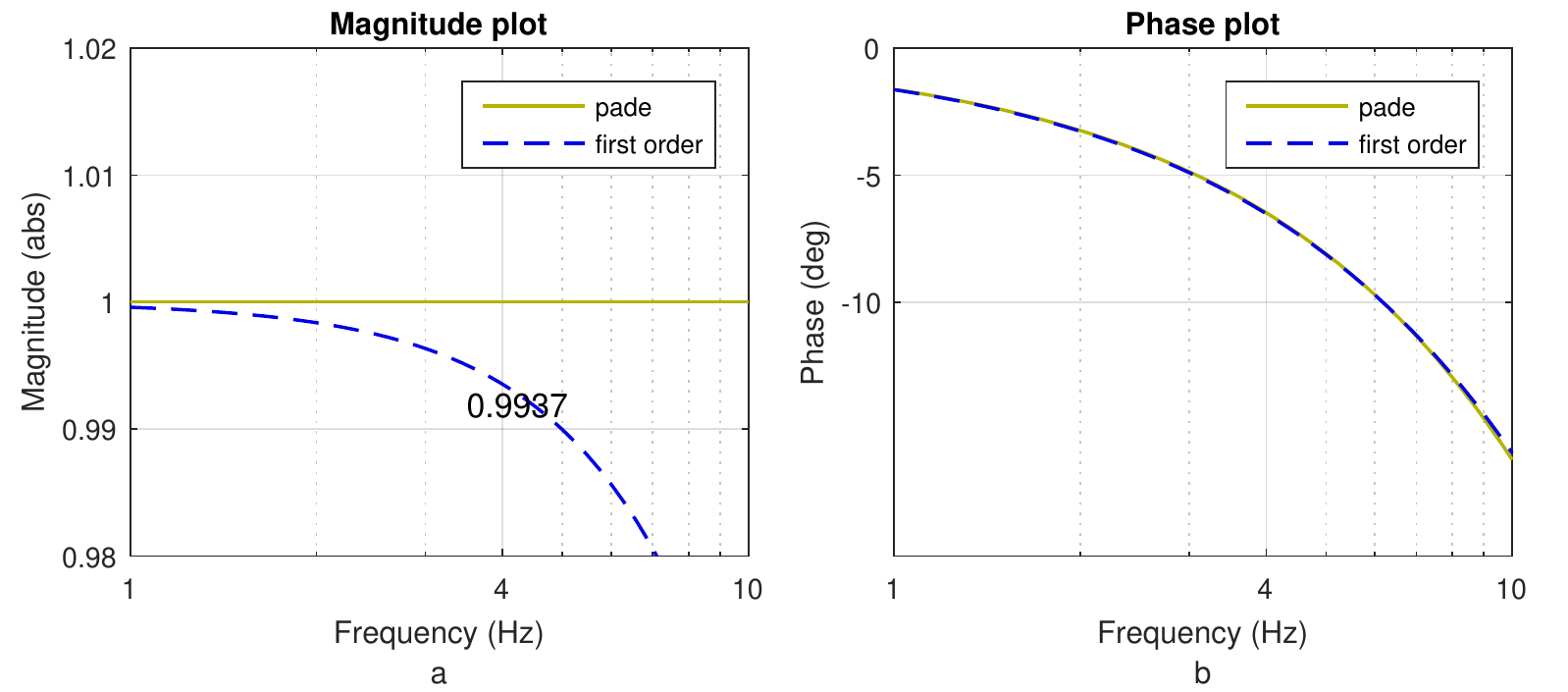}}
\end{figure}
Figure \ref{fi3} illustrates that approximation of second order Pade with a first order transfer function gives very small magnitude errors at 4 Hz by preserving the phase information. Since
the gimbal excitation is made with 4 Hz sinusoidal
signal, using first order low pass instead
of second order Pade gives very accurate result and saves a lot of computation effort \cite{baskin2015lqg}. So, the approximated transfer function and state space representation of the gimbal at 4 Hz
are given in \eqref{a13} and \eqref{a14}, respectively. Since $J$ and $B_v$ are unknown, they can be considered
as fifth and sixth state. This state augmentation leads to new nonlinear model in \eqref{a15}.
\begin{equation} \label{a13}
	G(s)\approx\frac{K_a K_t}
	{Js+B_v} \times \frac{w_g^2}{s^2+2\xi w_g s+w_g^2} \times \frac{1}{ds+1}	
\end{equation}
\begin{align} \label {a14}
	\begin{split}
		\dot{x}&= 
		\begin{bmatrix}
			0 & 1 & 0 & 0 \\
			-w_g^2 & -2\xi w_g & w_g^2 & 0 \\
			0 & 0 & -B_v/J & K_t/J \\
			0 & 0 & 0 & -1/d \end{bmatrix} x +
		\begin{bmatrix} 0 \\ 0  \\ 0 \\ 1/d \end{bmatrix}u \\
		y&=\begin{bmatrix} 1 & 0 & 0 &0 \end{bmatrix}x 
	\end{split}
\end{align}
\begin{equation} \label{a15}
	\dot{x}= 
	\begin{bmatrix}
		x_2 \\
		-w_g^2x_1-2\xi w_g x_2 +w_g^2 x_3\\
		(-x_6x_3+K_tx_4)/x_5\\
		(-x_4+u)/d \\
		0\\0\end{bmatrix}, y=x_1, x_5=J, x_6=B_v
\end{equation}
Observe that the gimbal model is continuous and the measurements are discrete. In this aspect, continuous-discrete EKF (CD-EKF) is considered. The detailed explanation of the CD-EKF can be found in \cite{baskin2015lqg,lewis2007optimal}. By using CD-EKF and model in \eqref{a15} the unknown
parameters are estimated, and online results are illustrated in Figure \ref{fi4}. The parameters in Table \ref{t2} is used for nominal plant construction.
\begin{table}
	\caption{Estimated parameters of the system  \label{t2}}
	\begin {center}
	\begin{tabular}{l l}
		\hline
		Parameters & Values \\
		\hline
		azimuth inertia, $J$ & $0.1736~kgm^2$ \\
		azimuth viscous friction, $B_v$ & $1.15~Nm/(rad/s)$ \\
		elevation inertia, $J$ & $0.063~kgm^2$ \\
		elevation viscous friction, $B_v$ & $0.61~Nm/(rad/s)$ \\
		\hline
	\end{tabular}
	\end {center}
\end{table}
\subsection{Nominal model construction}
The linearized two-axis gimbal can be represented with \eqref{a16} where $w_{az}$, $w_{el}$, $i_{az}$ and $i_{el}$ are the azimuth and elevation angular rates and current inputs to corresponding axes' motors.
\begin{equation} \label{a16}
	\begin{bmatrix} w_{az} \\ w_{el} \end{bmatrix}=
	\begin{bmatrix}
		G_{11} & G_{12} \\
		G_{21} & G_{22} \end{bmatrix}
	\begin{bmatrix} i_{az} \\ i_{el} \end{bmatrix}
\end{equation}
$G_{11}$ and $G_{22}$ are the transfer matrices of azimuth and elevation respectively having form
given in \eqref{a11} and parameters given in Table \ref{t1} and \ref{t2}.
 $G_{12}$ and $G_{21}$ transfer matrices are approximately zero when the gimbal is mass balanced. In the actual system, gains of these
transfer functions are indeed small and can be neglected \cite{baskin2015lqg}. In short, the MIMO nominal
model for two-axis gimbal is constructed and it will be used in the next sections.
\section{Design descriptions}
\subsection{Sensitivity weight selection}
Main objective of the LOS stabilization is to minimize the pointing error due to platform
motions. In electro-optical imaging systems root mean square (RMS) of the LOS error must
be smaller than the single detector pixel radiation angle for clear image acquisition. For that
reason, correct output sensitivity function which satisfies this constraint is found.
\begin{equation}\label{a17}
	S^{-1}=\frac{(s^2/M_s+2\xi w_bs/\sqrt{M_s}+w_b^2)}{(s^2+2\xi w_bs\sqrt{\epsilon}+w_b^2\epsilon)}, M_s=1,\epsilon=0.01,\xi=0.5,w_b=2\pi 10
\end{equation}
The sensitivity function given in \eqref{a17} gives approximately 75 microradian RMS LOS error under the known disturbance profile. Since this value is smaller than 100 microradian pixel radiation angle, the sensitivity in \eqref{a17} is good aim for stabilization loop design. Therefore, the sensitivity weight given in \eqref{a18} can be used for one axis. Similarly, for MIMO system, sensitivity weight in \eqref{a19} can be used in plant augmentation stage as
discussed in Section \ref{s2}.
\begin{equation}\label{a18}
	w_e=\frac{(s^2/M_s+2\xi w_bs/\sqrt{M_s}+w_b^2)}{(s^2+2\xi w_bs\sqrt{\epsilon}+w_b^2\epsilon)}, M_s=3.162,\epsilon=0.01,\xi=0.5,w_b=2\pi 10
\end{equation}
\begin{equation}\label{a19}
	W_e=\begin{bmatrix}
		w_e & 0 \\
		0 & w_e \end{bmatrix}
\end{equation}
\subsection{Uncertainty weight selection}
In this paper, output multiplicative uncertainty representation is used for model set representation. Firstly, the frequency response data of the gimbals are obtained by using swept sine tests at different excitation levels and at different gimbal positions. Next, using
the nominal model, possible multiplicative errors are found. Finally, the stable transfer functions which upper bound all these errors are obtained. The uncertainty upper bounds for azimuth and elevation axes in \eqref{a20} and \eqref{a21} are used while evaluating the robustness of the stabilization loop \cite{baskin2015lqg}. For MIMO system, the transfer matrix in \eqref{a22} is used.
\begin{equation}\label{a20}
	w_{1a}=\frac{1.87s^2+ 792.65s+90750}{s^2+ 650.35s+572624}
\end{equation}

\begin{equation}\label{a21}
	w_{1e}=\frac{1.12s^2+ 2564.28s+289957}{s^2+ 2059.65s+2375266}
\end{equation}

\begin{equation}\label{a22}
	W_e=\begin{bmatrix}
		w_{1a} & 0 \\
		0 & w_{1e} \end{bmatrix}
\end{equation}
By looking at the transfer functions \eqref{a20} and \eqref{a21}, it can be seen that at low frequencies
the uncertainties are around 0.15 and 0.12 for azimuth and elevation axes respectively. At
high frequencies due to the structural resonances of the gimbals, the uncertainties exceed 1
around 100 Hz and 200 Hz for azimuth and elevation axes respectively.
\section{LQG/LTR design}
As discussed in previous section, the transfer matrix $W_e$ which reflects the power spectrum of output disturbance $d$, is of order 4. Moreover, the nominal model constructed is of order 10. For that reason, the augmentation leads to generalized plant of order 14.
Hence the corresponding LQG/LTR controller will have an order of 14. Now, the LQG/LTR designs are investigated in detail.
\subsection{Design 1}
\begin{figure}[!hbt]
	\centering	
    \captionbox	
{Singular value plot of open loops and Kalman filter for design 1  \label{fi5}}
{\includegraphics[scale=0.65]{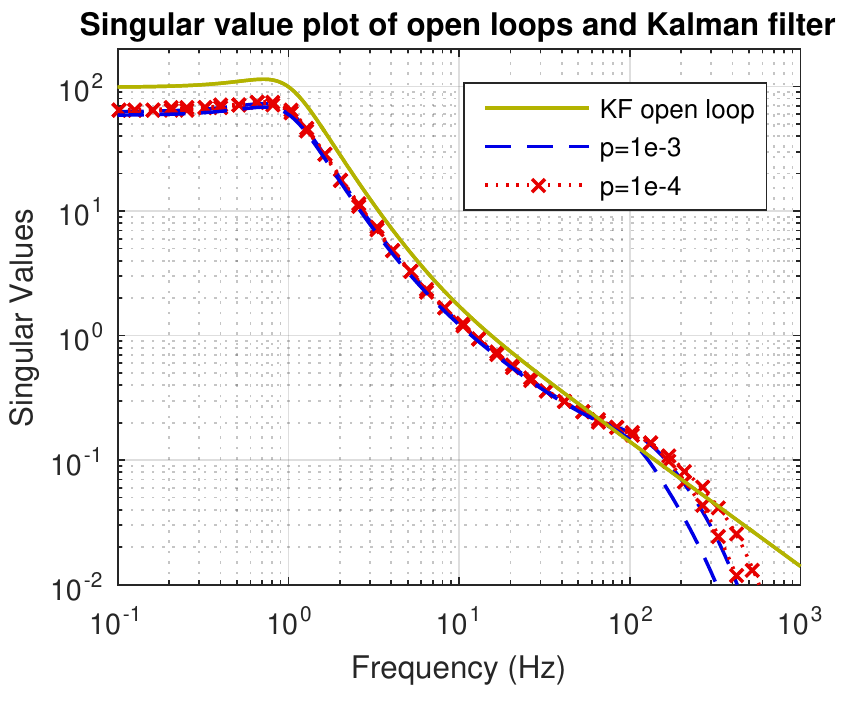}}
\end{figure}

The LQG/LTR controller is designed by using the procedure in Section \ref{s2}. As discussed previously, the aim is such that the open loop transfer matrix $GK_{LQG}$ needs to approach to
Kalman filter open loop transfer matrix $C\Phi K_f$. First, the Kalman filter is designed for augmented plant. Next, by reducing $\rho$, different optimal state feedbacks are designed, and the resulting open loop gains are given in Figure \ref{fi5}. As given in Figure \ref{fi5}, the recovery
procedure does not achieve the objectives successfully even if one continuously reduces $\rho$. The main result behind this fact is the non-minimum phase behavior of the gimbal. For successful loop recovery the plant zeros are usually canceled by the compensator poles \cite{maciejowski1989multivariable}.
Since this is not possible for non-minimum phase plants, the procedure success reduces. To get rid of this drawback, the design 1 is reconsidered and design 2 is made.
\subsection{Design 2}
\begin{figure}[!hbt]
	\centering
	\captionbox	
	{Singular value plot of design 2 \\ a Singular value plot of Kalman filters and $W_e$ for design 2\\ b Singular value plot of open loops and Kalman filter for design 2 \label{fi6}}
	{\includegraphics[scale=0.7]{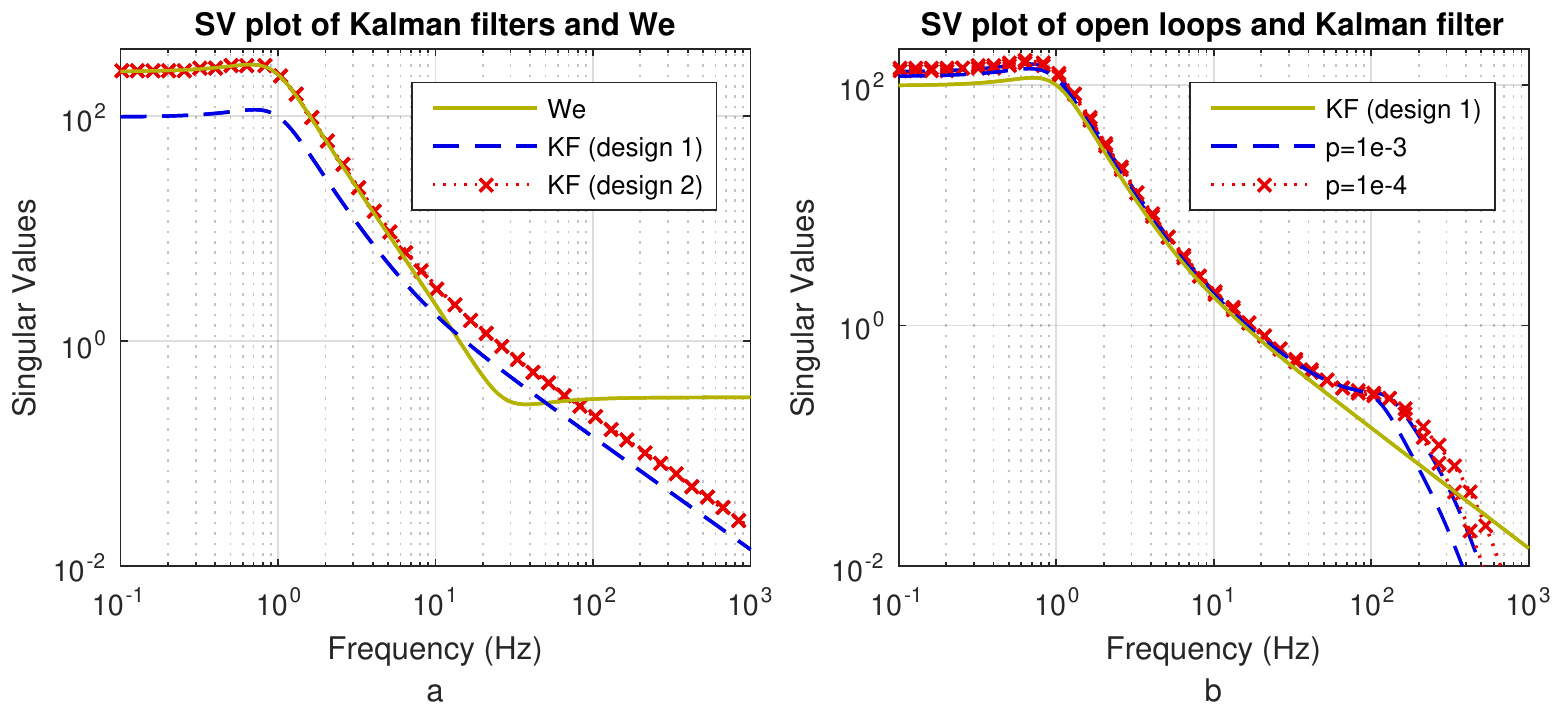}}
\end{figure}

For design 2, the transfer matrix $W_e$ is modified such that it has a higher bandwidth and a
dc gain. For this case, the Kalman filter from two designs and modified weight $W_e$ is plotted
in Figure \ref{fi6}a. Then different optimal state feedbacks are designed by reducing $\rho$, and the resulting open loop singular values are given in Figure \ref{fi6}b.\\
\indent Observe that, for non-minimum phase plants exact recovery is not possible. However, by augmenting the plant with a new weight and demanding more performance, the design that is better than design 1 can be obtained. In other words, while trying to recover Kalman filter of design 2, it is possible to recover Kalman filter of design 1 approximately. As given in Figure \ref{fi6}b, the recovery procedure is made such that the objectives of the design 1 are recovered. To do that, the weighting matrix is modified such that the bandwidth is enlarged to 15 Hz from 10 Hz, and 2.5 multiples of the dc gain is used.
\section{Robustness analysis with singular value tests}
\begin{figure}[!hbt]
	\centering	
\captionbox	
{Performances and stabilities \\a Nominal performances \\b Robust performances and robust stabilities \label{fi8}}
{\includegraphics[scale=0.7]{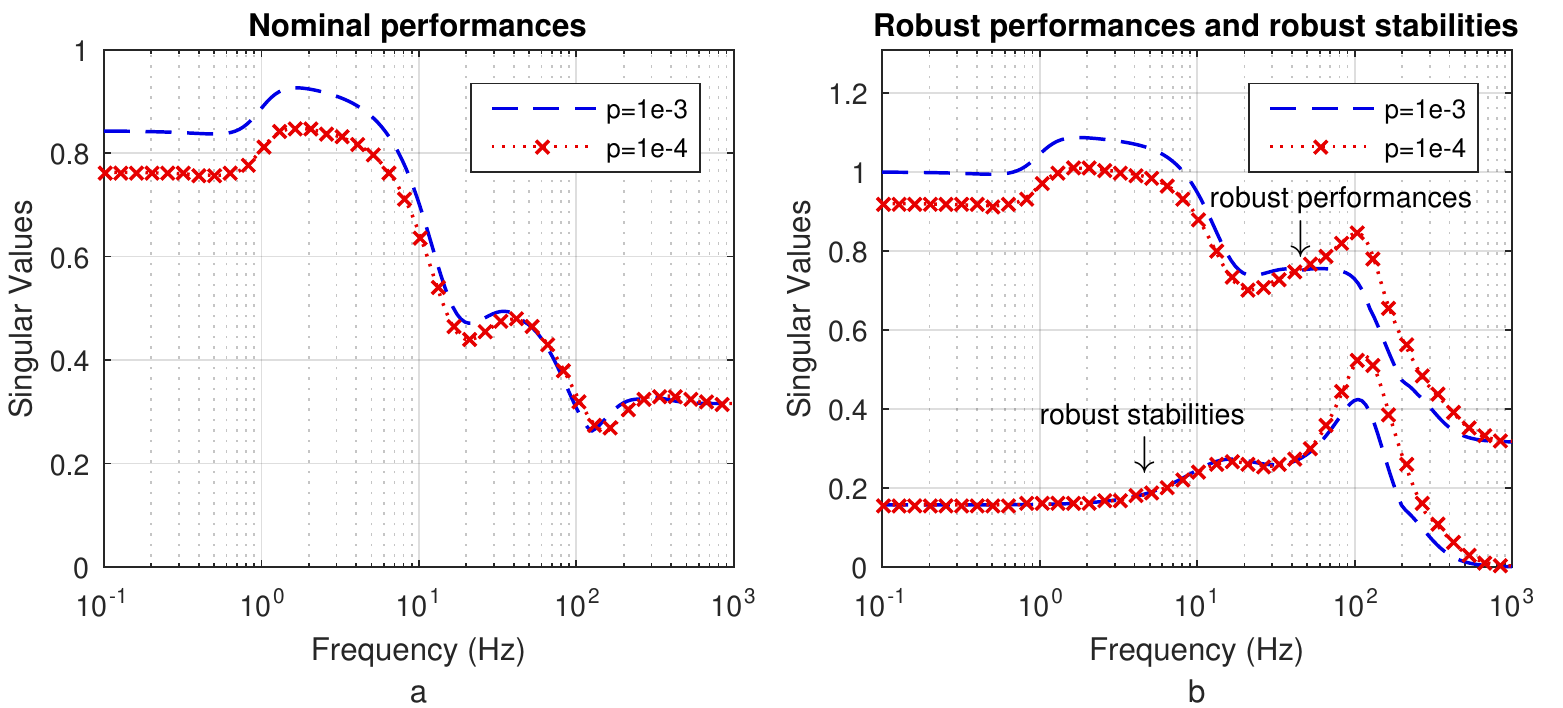} }	
\end{figure}

The aim of the stabilization loop is to satisfy disturbance rejection constraint for all models in the output multiplicative model set. 
\begin{figure}[!hbt]
	\centering	
	\captionbox	
	{Structure for robustness analysis  \label{fi7}}
	{\includegraphics[scale=0.7]{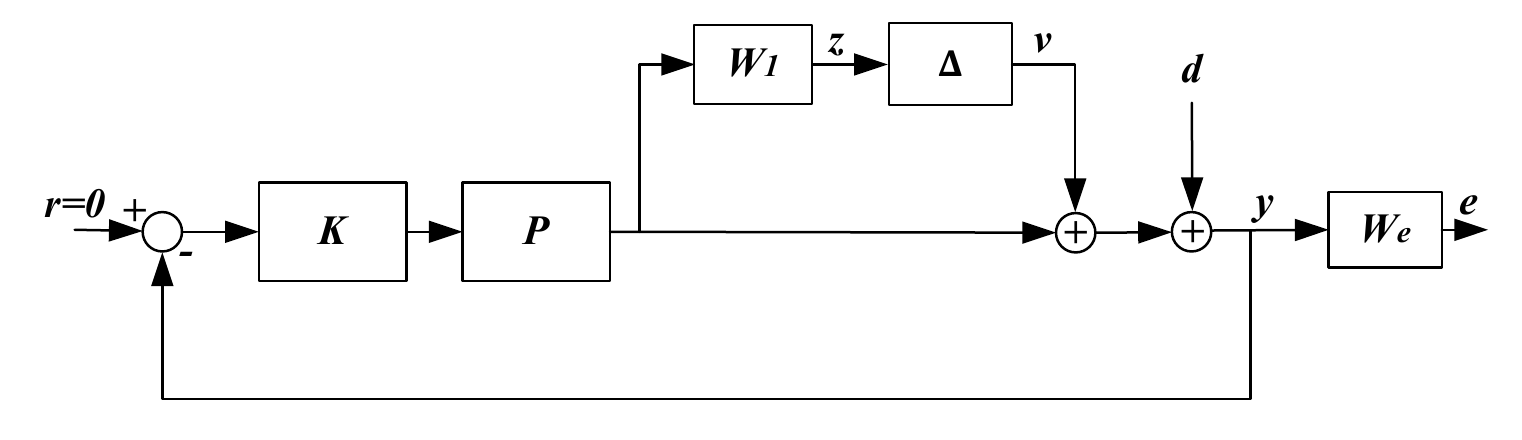}}
\end{figure}
For that reason, the structure in Figure \ref{fi7} is used for robustness analysis. Figure \ref{fi7} shows that performance index is measured at plant
output and there is only one uncertainty block in the structure. For this special structure, the robustness can be evaluated using just singular value tests, and there is no need to investigate structured singular value $(\mu)$. The nominal performance, robust stability and
robust performance tests are given in \eqref{a23} to \eqref{a25} respectively \cite{zhou1998essentials,skogestad2007multivariable}.
\begin{equation} \label{a23}
	\lVert W_eS_o\rVert _\infty <1
\end{equation}
\begin{equation} \label{a24}
	\lVert W_1T_o\rVert _\infty <1
\end{equation} 
\begin{equation} \label{a25}
	\bar{\sigma}(W_eS_o)+\bar{\sigma}(W_1T_o) <1,\forall w
\end{equation}
Observe that the structure is very special such that the robust performance test is just an addition of nominal performance and robust stability tests. Since the recovery is not
satisfied for design 1, these tests are applied only to design 2. Figure \ref{fi8} shows that with
controllers in design 2, nominal performance and robust stability are satisfied for both $\rho$
values. Since the peak value of robust performance test is very close to 1, robust performance can be assumed to be satisfied for $\rho = 1e^{-4} $. For $\rho=1e^{-3}$ the robust performance
is not satisfied; however, it leads to more stable loop. In short, reducing $\rho$ makes the
performance better at the cost of reducing stability. This situation results from the high
controller gains at high frequencies for small $\rho$ value.
\section{Implementation}
The LQG/LTR controller obtained with $\rho = 1e^{-4}$ is selected since it approximately
satisfies the robust performance. This controller is of order 14. To reduce the process cost of
the implementation, reduced order controller is obtained with Balanced Truncation method.
The reduced order controller is of order 12 and does not yield any performance degradation
\cite{baskin2015lqg}. Next, this reduced order controller is discretized with bilinear transform and implemented in digital computer.
\section{Experimental Results}
\begin{figure}[!hbt]
	\centering
	\captionbox	
	{Experimental results\\ a Singular value plot of sensitivities and $W_e^{-1}$ \\ b Performances of perturbations \label{fi9}}
	{\includegraphics[scale=0.7]{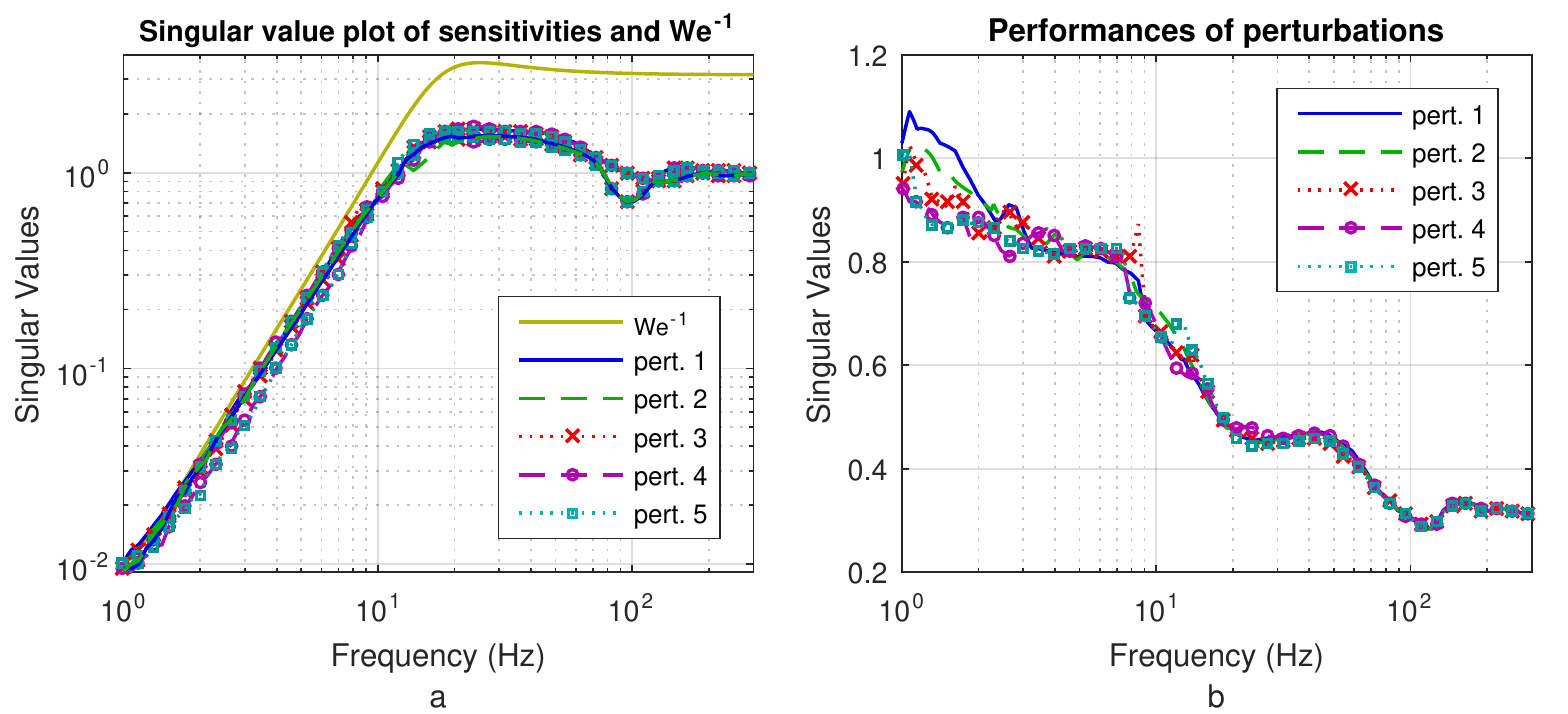}}	
\end{figure}

The equation of the classical closed loop system is given in \eqref{a26}. If reference $r$, noise $n$
and input disturbance $d_i$ are assumed to be zero, simple equation in \eqref{a27} is obtained for a two-axis gimbal.
\begin{equation} \label{a26}
	y=T_o(r-n)+S_oPd_i+S_od
\end{equation}
\begin{equation} \label{a27}
	\begin{bmatrix} w_{az} \\ w_{el} \end{bmatrix}=
	\begin{bmatrix}
		S_{o11} & S_{012} \\
		S_{o21} & S_{o22} \end{bmatrix}
	\begin{bmatrix} d_{az} \\ d_{el} \end{bmatrix}
\end{equation}
Using discretized reduced order controller, the closed loop is constructed. Next, closed
loop sensitivity responses are obtained by using swept sine tests. By making $d_{el}$ zero, $S_{o11}$
and $S_{o21}$ are determined by looking at $w_{az}$ and $w_{el}$ respectively. Similarly, under zero $d_{az}$,
$S_{o12}$ and $S_{o22}$ are found. 
After finding responses of corresponding transfer functions, for two-input two-output system transfer matrix is constructed. 
Then the singular values of the sensitivity matrix are obtained and shown in Figure \ref{fi9}a. 
After that, the performance $\lVert W_eS_o\rVert _\infty$ is evaluated and illustrated in Figure \ref{fi9}b for different model perturbations.\\
\indent The theoretical performances were given in Figure \ref{fi8} before. The experimental results
possess similar characteristics. 
Figure \ref{fi9} shows that the sensitivity is successfully shaped and the robust performance is approximately satisfied.
\section{Conclusion}
The theoretical and experimental results show that with the introduced LQG/LTR
method the sensitivity shaping is simple and efficient. Moreover, the designed closed loop
gives good results for both nominal model and any model in the model set.
All results show that when the performance is measured only by sensitivity, LQG/LTR method can satisfy
the robust performance. 
However, when more sophisticated performance index is available,
LQG/LTR method can be insufficient \cite{baskin2015lqg}.
\section{Acknowledgments}
The support and facilities provided by ASELSAN Inc. are gracefully appreciated.
\bibliographystyle{IEEEtran}
\bibliography{bibl}
\end{document}